\documentclass{aip-cp}

\usepackage[numbers]{natbib}
\usepackage{rotating}
\usepackage{graphicx}
\usepackage{url}
\usepackage[utf8]{inputenc}

\begin{document}

\title{gamma-sky.net: Portal to the Gamma-Ray Sky}
\corresp[cor1]{Corresponding author: arjun.voruganti@gmail.com}

\author[mpik]{Arjun Voruganti\corref{cor1}}
\author[mpik]{Christoph Deil}
\author[mpik]{Axel Donath}
\author[mpik]{Johannes King}

\affil[mpik]{MPIK, Heidelberg, Germany}

\newcommand{\gammasky}{\texttt{gamma-sky.net} }
\newcommand{\gammaskyurl}{\url{http://gamma-sky.net/} }
\newcommand{\gammaskygh}{\url{https://github.com/gammapy/gamma-sky} }
\newcommand{\gammacat}{\url{https://github.com/gammapy/gamma-cat} }

\maketitle

\begin{abstract}
\url{http://gamma-sky.net} is a novel interactive website designed for exploring the gamma-ray sky. The Map View portion of the site is powered by the Aladin Lite sky atlas, providing a scalable survey image tesselated onto a three-dimensional sphere. The map allows for interactive pan and zoom navigation as well as search queries by sky position or object name. The default image overlay shows the gamma-ray sky observed by the Fermi-LAT gamma-ray space telescope. Other survey images (e.g. Planck microwave images in low/high frequency bands, ROSAT X-ray image) are available for comparison with the gamma-ray data.
Sources from major gamma-ray source catalogs of interest (Fermi-LAT 2FHL, 3FGL and a TeV source catalog) are overlaid over the sky map as markers. Clicking on a given source shows basic information in a popup, and detailed pages for every source are available via the Catalog View component of the website, including information such as source classification, spectrum and light-curve plots, and literature references.

We intend for \gammasky to be applicable for both professional astronomers as well as the general public. The website started in early June 2016 and is being developed as an open-source, open data project on GitHub (\gammaskygh). We plan to extend it to display more gamma-ray and multi-wavelength data. Feedback and contributions are very welcome!

\end{abstract}

\section{Introduction}

The field of gamma-ray astronomy is growing tremendously – while only a decade ago no more than a handful of sources were observed in the GeV range, today thousands have been found, including hundreds within the TeV range. This advancement has been made possible due to novel ground-based Cherenkov telescope instruments for very-high-energy (VHE; $>$ 100 GeV) source detections. Such systems exhibit more accurate detections and higher angular resolutions than ever before. Space-based satellites sharing similar technological breakthroughs have further developed the high-energy (HE; 10 MeV -- 100 GeV) range of gamma-ray astronomy, as can be observed in the latest images from the Fermi Large Area Telescope (Fermi-LAT). As a whole, the instruments can capture gamma-rays in a wide spectrum of energies from 10 MeV to 10 TeV. The High Energy Stereoscopic System (H.E.S.S.) Galactic Plane Survey \cite{hgps}, the High-Altitude Water Cherenkov Observatory (HAWC) 1st Year Catalog, and the fourth Fermi-LAT Point Source Catalog (4FGL) are among the highly anticipated surveys that will be unveiled in the near future. Furthermore, with an incoming wave of notable systems planned to operate soon, such as the ground-based Cherenkov Telescope Array (CTA) for TeV observations \cite{cta}, numerous never-before-seen sources in the gamma-ray sky are expected to be discovered. With such abundance of HE and VHE sources and a rapid growth of interest in this field, there is an increasingly evident need for a central hub for exploring relevant catalog and image data across the gamma-ray sky. Our website (\url{http://gamma-sky.net}) was designed to function as such.

\section{Idea}

\gammasky is a one-stop resource for browsing images and catalogs but also for closely examining a specific gamma-ray source. Although it was mainly built for the greater astronomical community, the web page additionally targets the general public through a user-friendly interface and a clean information layout, all of which are compiled under cutting-edge web tools.

\begin{figure}[tb]
\centerline{\includegraphics[width=\textwidth]{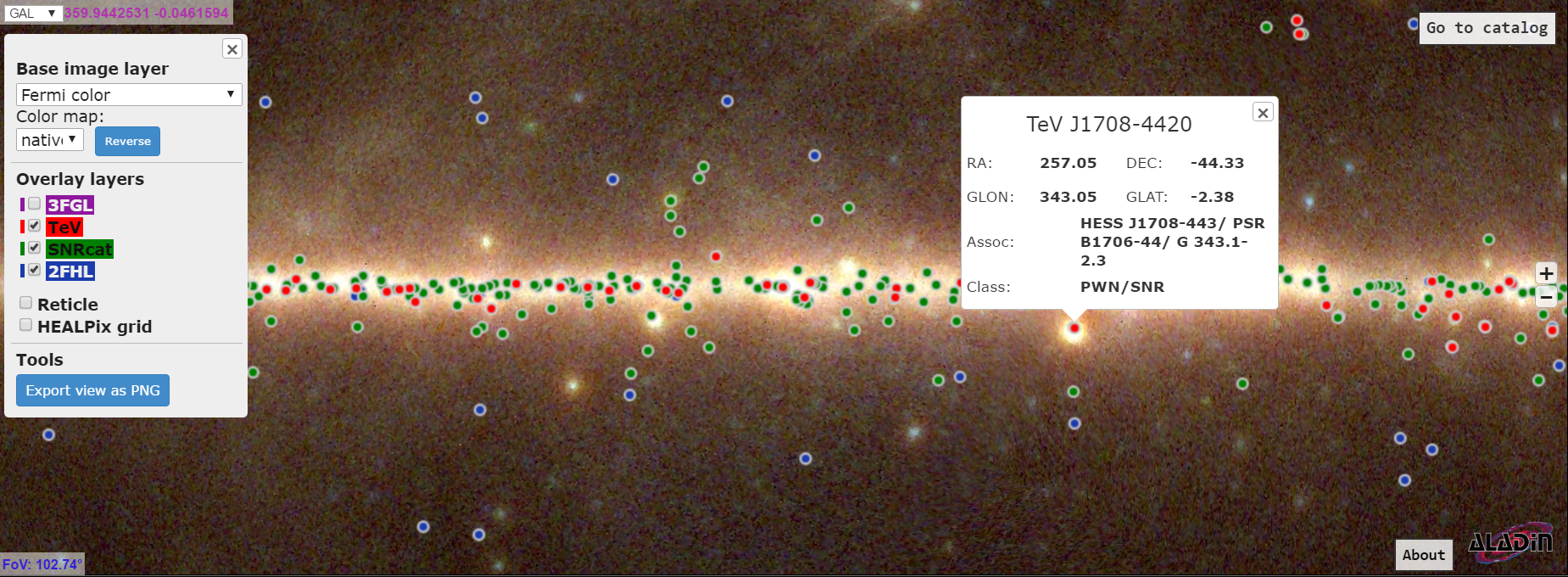}}
\caption{Map View of \gammasky showing the gamma-ray sky observed by the space-based Fermi-LAT, broken into three bands: 0.3--1 GeV (red-band), 1--3 GeV (green-band), and 3--300 GeV (blue-band). The pop-up for one TeV source is displayed, as well as the toolbar on the side containing widgets for controlling the map overlays.}
\label{fig:mapview}
\end{figure}

\begin{figure}[tb]
\centerline{\includegraphics[width=\textwidth]{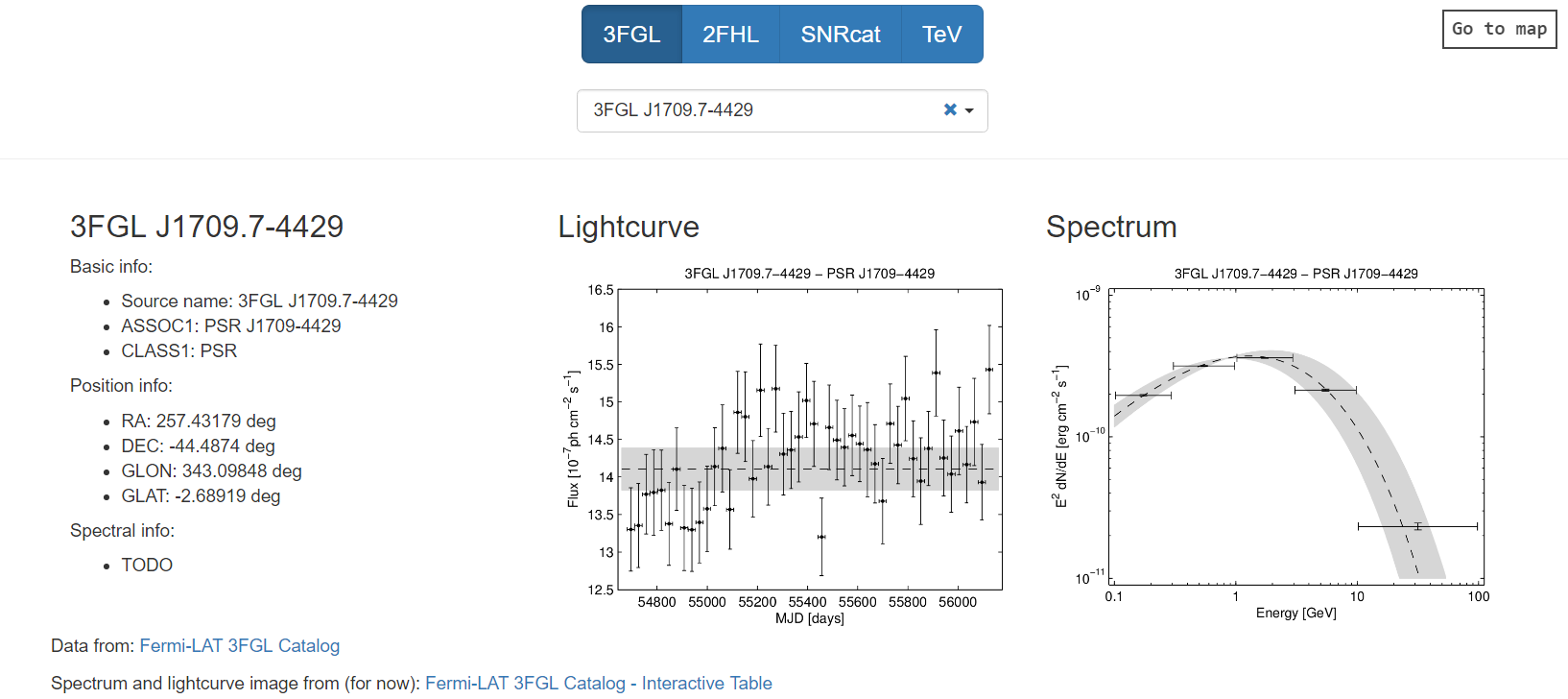}}
\caption{Catalog View of \gammasky showing the catalog and source selection fields at the top, and the detailed view for one source from the Fermi-LAT 3FGL catalog.}
\label{fig:catview}
\end{figure}

Individuals who access \url{http://gamma-sky.net} via any modern internet browser will be welcomed with the Map View homepage. This page displays gamma-ray and multi-wavelength survey images, most of which are all-sky, wrapped around a three-dimensional sphere. Gamma-ray sources from our catalog data have then been pinpointed onto the sphere, as shown in Figure~\ref{fig:mapview}. The map features pan-and-zoom functionality for easily navigating and quickly browsing the sky. The Map View page also utilizes a powerful search tool to either pan the view to a given sky position or locate a source by name. This functionality is intended to work similar to Google Maps\footnote[1]{https://www.google.com/maps} and allows the user to easily find sources of interest and study their visual context. See Table~\ref{tab:images} for our selection of images on the website and Table~\ref{tab:catalogs} for the catalogs we display. \gammasky additionally embodies a Catalog View, which incorporates more detailed information for each of the sources in our catalogs. Professional astronomers will navigate to this component of the website for the deep investigation of a particular source. The Catalog View page is shown in Figure~\ref{fig:catview}.

\begin{table}[bt]

\caption{
Survey images of interest to gamma-ray astronomers available on the Aladin Lite map at \gammasky . Note that these surveys only make up very small selection of the 300+ multi-wavelength survey images available from CDS' HiPS database (\protect\url{http://aladin.u-strasbg.fr/hips/list}). We plan to add gamma-ray survey images in the GeV energy range (high-energy Fermi-LAT data) and the TeV energy range (H.E.S.S. Galactic Plane Survey) to the website when they become available.
}

\label{tab:images}
\tabcolsep7pt\begin{tabular}{ lll }
\hline
Image              & Waveband  & Description\\
\hline
Fermi color        & gamma-ray & Fermi-LAT RGB all-sky survey (0.3--1 GeV, 1--3 GeV, 3--300 GeV) \\
RASS               & x-ray & ROSAT all-sky survey (0.1--2.4 KeV) @ 1.8 arcmin\\
XMM-EPIC-RGB       & x-ray & XMM-Newton EPIC RGB composition\\
AKARI 90um         & infrared  & AKARI Wide-S all-sky survey @ 1 arcmin\\
Spitzer GLIMPSE360 & infrared  & Spitzer galactic plane survey @ 0.02 arcmin\\
IRIS Band 4-100um  & infrared  & IRIS all-sky survey\\
Planck R1 + R2 HFI & microwave & Planck 353-545-857 GHz all-sky survey\\
Planck R2 LFI      & microwave & Planck 30-44-70 GHz all-sky survey\\
CGPS-VGPS CONT     & radio     & galactic plane continuum @ 1 arcmin\\
Haslam 408         & radio     & 408 MHz all-sky continuum @ 51 arcmin\\
\hline
\end{tabular}

\end{table}

\begin{table}[bt]

\caption{
Source catalogs currently displayed on \gammasky .
The \texttt{gamma-cat} and \texttt{SNRcat} catalogs are live and updated often,
while the Fermi catalogs are fixed and never updated.
We intend to add additional catalogs of interest to gamma-ray astronomers to the website in the future, including the upcoming H.E.S.S. and HAWC TeV source catalogs, as well as the ATNF Pulsar Catalogue (\protect\url{http://www.atnf.csiro.au/people/pulsar/psrcat/}).
}
\label{tab:catalogs}
\tabcolsep7pt\begin{tabular}{ lrll }
\hline
Catalog   & Sources & Type    & Description \\
\hline
TeV &     163 & live    & gamma-cat: Open TeV gamma-ray source catalog  \\
&&& \gammacat  \\
2FHL      &     360 & fixed   & Second Fermi-LAT catalog of high-energy sources \citep{2fhl}\\
&&& \url{http://fermi.gsfc.nasa.gov/ssc/data/access/lat/2FHL/}  \\
3FGL      &    3034 & fixed   & Third Fermi-LAT point source catalog \citep{3fgl}\\
&&& \url{http://fermi.gsfc.nasa.gov/ssc/data/access/lat/4yr_catalog/}  \\
SNRcat    &     378 & live    & A census of high-energy observations of Galactic supernova remnants \citep{snrcat}\\
&&& \url{http://www.physics.umanitoba.ca/snr/SNRcat/} \\
\hline
\end{tabular}
\end{table}

\begin{figure}[tb]
  \centerline{\includegraphics[width=\textwidth]{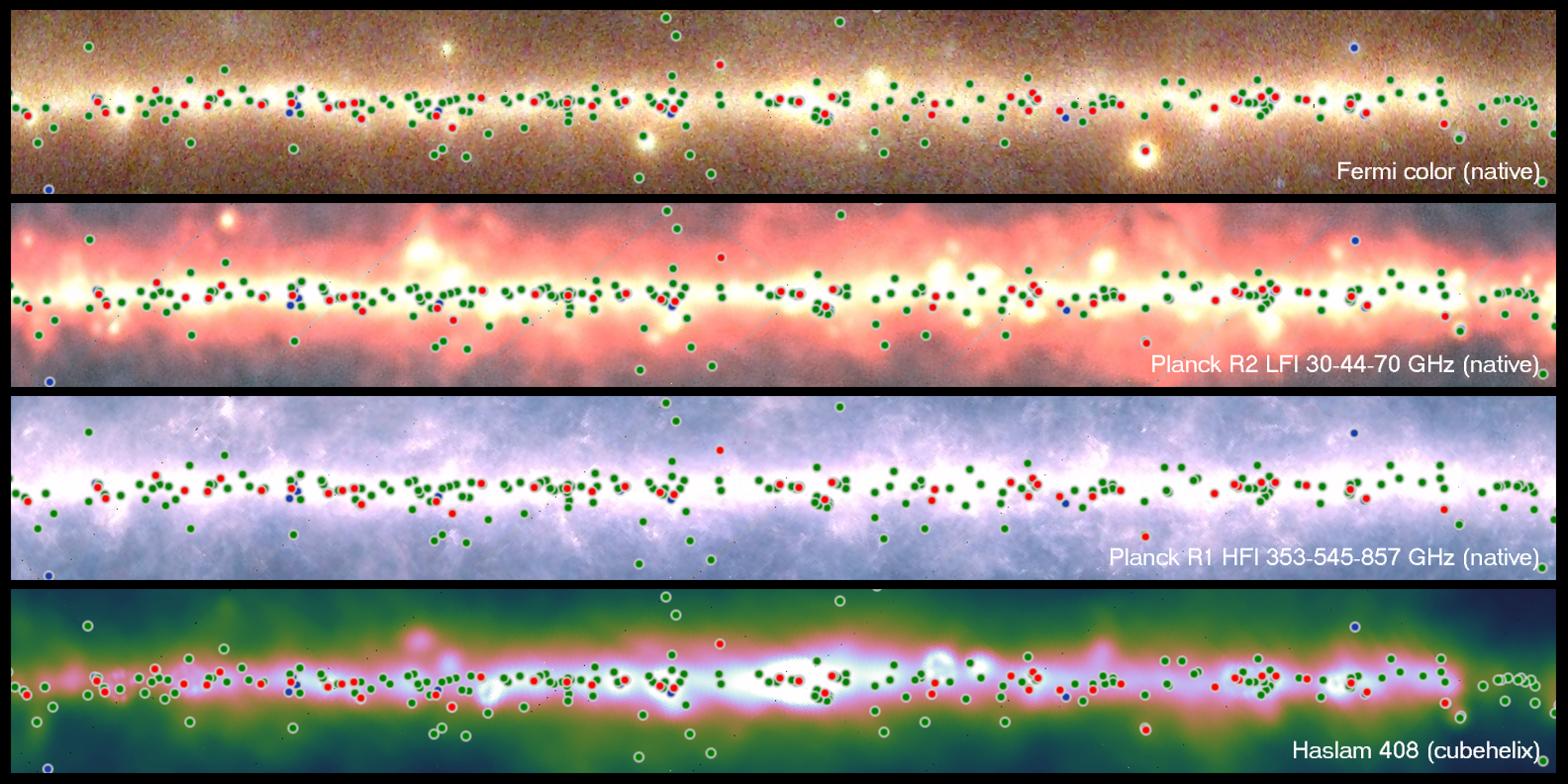}}
  \caption{The inner galaxy in various survey images and color maps, FOV $90 \times 15$ degrees. The red (TeV) and green (SNRcat) markers indicate many galactic supernova remnants (SNRs) and pulsars detected at very-high-energies.}
  \label{fig:galactic}
\end{figure}

\begin{figure}[tb]
  \centerline{\includegraphics[width=\textwidth]{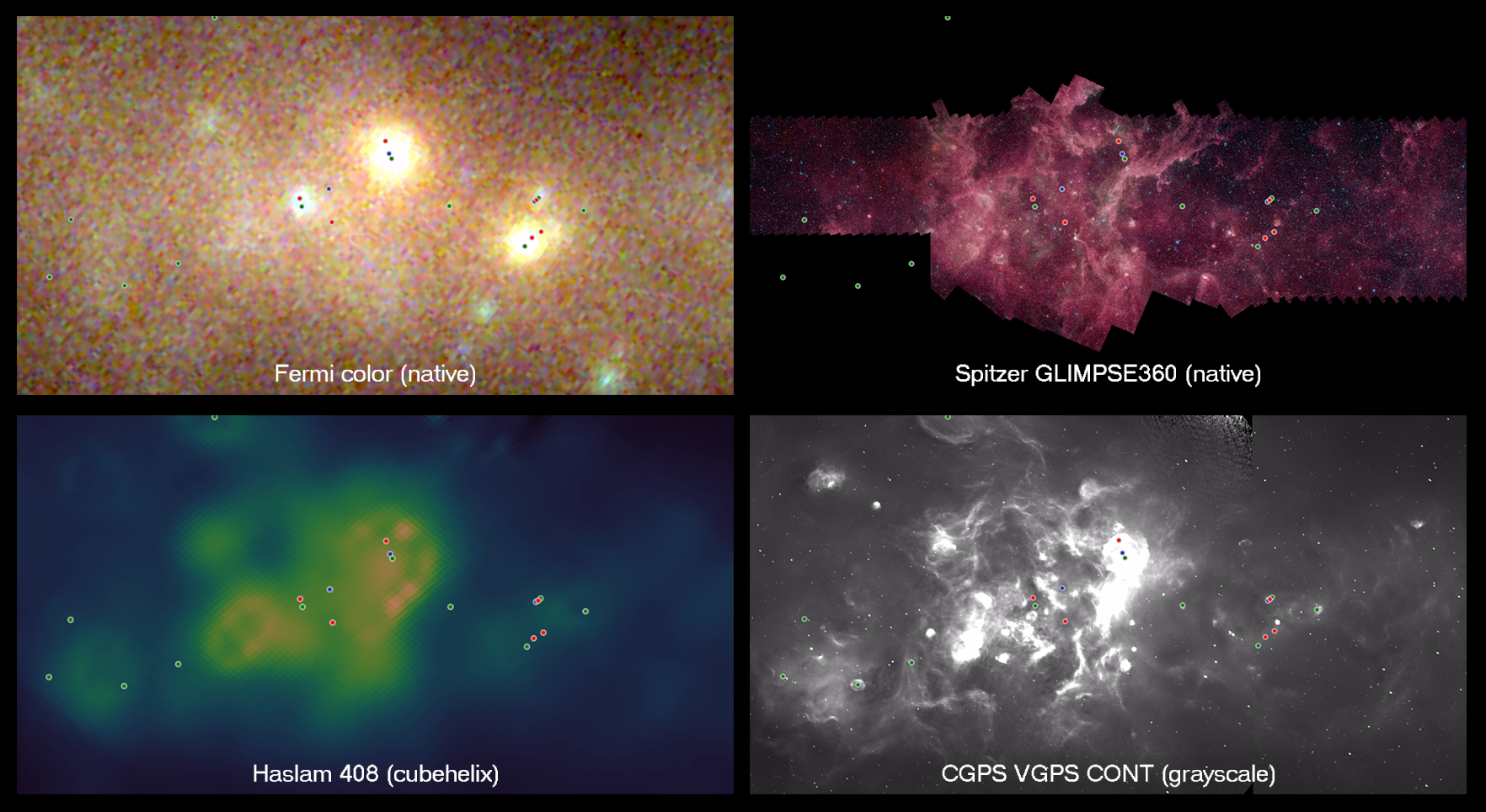}}
  \caption{The Cygnus Region in various survey images and color maps, FOV $20 \times 10$ degrees. Sources of interest include: the diffuse emission nebula surrounding the Gamma Cygni star, the cosmic-ray Cygnus Cocoon observed by Fermi, and a handful of unidentified TeV sources indicated with red (TeV) markers.}
  \label{fig:cygnus}
\end{figure}

\gammasky aims to be a resource for browsing a region or examining a specific source (see Figures~\ref{fig:galactic}~and~\ref{fig:cygnus} for examples); however, any tools for extended analysis will not be incorporated into the website. Instead of limiting analysis to only pre-built functionality facilitated through the web page, we want to allow users to be able to utilize data for their specific use case. We point users to Gammapy \cite{gammapy}, a Python package for gamma-ray astronomy, for continued investigation at the local level. Gammapy's scripts will generate the detailed plots and publication-worthy results that astronomers are interested in. Alternatively, users may navigate to TeVCat \cite{tevcat} for investigating TeV sources, or the ASI Science Data Center (ASDC) website for their online tools and TeGeV Catalogue for TeV sources \cite{tgevcat}.

The motivation for creating \gammasky was to provide a resource for quickly browsing a source or region in the gamma-ray sky. We present our catalog and image data in a clean view, utilizing modern web development tools, for finding source information with ease and minimal effort. These survey images and catalogs we selected are greatly of interest to the gamma-ray astronomical community. It is imperative for all of our data to be openly available for download so that astronomers can produce their own results in extended local analysis. Additionally, the website is an entirely open-source project, and other developers are welcome to contribute to the code. We advise those interested in contributing to visit our GitHub repository at \url{https://github.com/gammapy/gamma-sky}.

\renewcommand{\thefootnote}{\fnsymbol{footnote}}

\section{Features}

\begin{enumerate}

\item \textbf{Map View} interface from the Aladin Lite tool \cite{aladin-lite}:

\begin{itemize}
\item Pan and zoom
\item Sesame search term resolver - locate objects by name, association, or coordinate position
\item Toggle and view specific catalog layers and sky images
\item Pop-up information over each source
\item Export and share images from the sky map (in PNG format)
\end{itemize}

\item \textbf{Catalog View} and its information panels for a given source:

\begin{itemize}

\item Search and select a source by its name
\item Basic information - position, association and class
\item Extension information
\item Spectral index, brightness and flux
\item Distance and redshift
\item Light curve and emission spectrum plots
\item Detection/observation information - instrument, date of discovery and literature references

\end{itemize}

\end{enumerate}

\renewcommand{\thefootnote}{\arabic{footnote}}

\section{Data}

\begin{figure}[tb]
  \centerline{\includegraphics[width=\textwidth]{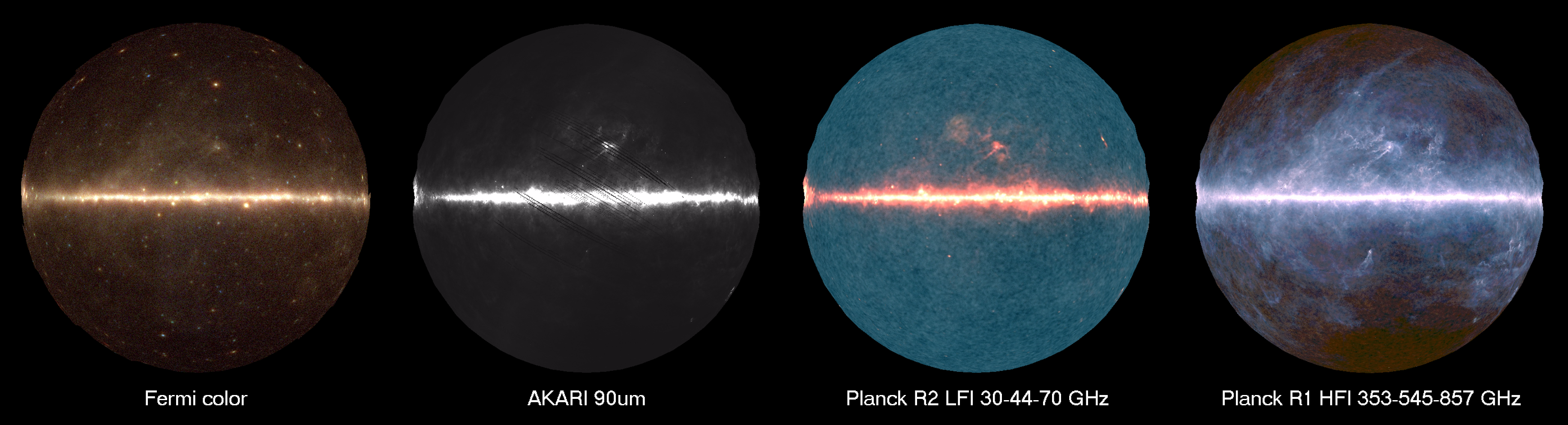}}
  \caption{Survey images (left to right): Fermi color, AKARI 90um, Planck LFI, Planck HFI. Images centered on the Galactic Center, FOV 180 degrees.}
  \label{fig:four_images}
\end{figure}



%

The default image layer displayed on the website's Map View page is a multi-wavelength all-sky survey from Fermi-LAT. As Fermi data is broken into multiple bands of energies, the map's native color scheme represents detected energies as categorized colors - red/yellow for the 0.3--1 GeV band, green for the 1--3 GeV band, and blue for the 3--300 GeV band.

 The Fermi color image is presented on the website as a Hierarchical Progressive Survey (HiPS) image \cite{hips}. HiPS is a hierarchical data structure utilizing the HEALPix\footnote[2]{\url{http://healpix.sourceforge.net/}} tesselation of a sphere that organizes data onto pixelated tiles of scalable resolution. The image mechanism allows catalog data and source markers on \gammasky to be visualized accurately on the sky map at various zoom levels. The Centre de Donn\'{e}es astronomiques de Strasbourg (CDS) developed the HiPS technology, and \gammasky currently encompasses 10 survey images also prepared by CDS in this format. The 10 images, which are outlined in Table~\ref{tab:images}, come from CDS's HiPS database\footnote[3]{\url{http://aladin.u-strasbg.fr/hips/list}} of over 300 prepared HiPS images. Four selected images are displayed on the three-dimensional sphere in Figure~\ref{fig:four_images}.

Our website incorporates 4 source catalogs, as illustrated in Table~\ref{tab:catalogs}. 3FGL \cite{3fgl} and 2FHL \cite{2fhl} are the latest source catalogs from Fermi-LAT, the main space-based instrument we display sources from. SNRcat \cite{snrcat} is an up-to-date compilation of galactic supernova remnants (SNRs) observed from a variety of instruments. The database is maintained by Gilles Ferrand and Samar Safi-Harb. gamma-cat is an open-data catalog of sources in the TeV range. As a project that has just recently begun in early September 2016, it is undergoing rapid growth and will be updated frequently on \gammasky. gamma-cat was started at the Max-Planck-Institut f\"{u}r Kernphysik (MPIK) and is open to contribution from other developers.

User inputs for search fields under the Map View portion of the website are interpreted by the Sesame service\footnote[4]{\url{http://cds.u-strasbg.fr/cgi-bin/Sesame}}. Sesame is a search term resolver for astronomical objects which queries several databases and returns the resolved sources. Both Sesame and the databases searched (SIMBAD, NED, and VizieR) are maintained by CDS.


\section{Implementation}

Scientific Python packages Astropy \cite{astropy} and Gammapy were used to generate all of the catalog and source data on \gammasky. This step includes preparing formatted JSON files from fetched raw FITS data for display in our Catalog View panels. Serving our catalog data in such manner requires us to frequently update the ``live" catalogs SNRcat and gamma-cat stored in our server. All data is consumed with the JavaScript and HTML front-end. The website's architecture was organized using Angular 2\footnote[5]{\url{https://angular.io/}}, a modern web application framework for JavaScript. Using Angular 2 has allowed us to compile the site into a single-page application. The sphere interface and visualization in the Map View page was implemented under the Aladin Lite tool \cite{aladin-lite} developed at CDS. The website is being hosted by GitHub Pages.

\gammasky exists as a static web page, meaning that there is no back-end server doing real-time data processing or content generating on request. The JavaScript client is still capable of downloading any static assets, including very large HiPS survey images and catalog datasets. Advantages to the static website technology include easy implementation and maintenance, as well as efficient data loading within the GitHub Pages server.

\section{Status and Outlook}

Our website is a new project, having been deployed very recently at \url{http://gamma-sky.net} in early June 2016. The current content of \gammasky is simply a starting point; we have plans to greatly expand on our catalog and image data. Such data includes additional image surveys from CDS' HiPS database, as well as data from upcoming surveys upon their public release. The Fermi high-energy images are among the images we will prepare with HiPS for display on the Map View. We additionally strive to enhance the user interface of \gammasky through additional features, including new source groupings by classification as done in NASA's Fermi-LAT 3FGL Catalog Interactive Table\footnote[6]{\url{http://fermi.gsfc.nasa.gov/ssc/data/access/lat/4yr_catalog/3FGL-table}} and more intricate data panels for the Catalog View. Using Angular 2, an improved routing network will be implemented  to allow for sharing a specific view in the Map View page by URL. We will continue to point directly to Gammapy scripts for any further analysis and keep server-heavy tools off our website.

\section{Acknowledgements}

We would like to thank CDS (Centre de Donn\'{e}es astronomiques de Strasbourg) for developing the data formats (HiPS) and tools (Aladin Lite) that make \gammasky possible. Thomas~Boch helped us configure Aladin Lite to our use case. We would also like to thank: GitHub for hosting our website; CDS for serving the HiPS images; and all of the contributors to the open-source projects, including Angular 2, Astropy, and Gammapy, that we have used to build \gammasky.


\nocite{*}
\bibliographystyle{aipnum-cp}
\bibliography{gammaskynet-gamma2016}

\end{document}